\documentclass{PoS}

\newcommand{\etal}{\textit{et al.}}

\newcommand{\eg}{{\sl e.g. }}

\newcommand{\gev}{{\ensuremath\rm GeV}}

\newcommand{\drbar}{{\ensuremath\mathrm{\overline{DR}}}}
\newcommand{\mRe}{{\ensuremath\mathrm{Re}}}
\newcommand{\di}{{\ensuremath\mathrm{d}}}
\newcommand{\om}{\ensuremath\mathcal{O}}

\title{Supersymmetric Higgs Production in Vector-Boson Fusion}

\ShortTitle{Supersymmetric Higgs Production in Vector-Boson Fusion}

\author{\speaker{Michael Rauch}\\
        Institut f\"ur Theoretische Physik, 
        Karlsruhe Institute of Technology (KIT)\\
        E-mail: \email{rauch@particle.uni-karlsruhe.de}}

\author{Wolfgang Hollik\\
        Max-Planck-Institut f\"ur Physik, M\"unchen\\
        E-mail: \email{hollik@mppmu.mpg.de}}

\author{Tilman Plehn\\
        Institut f\"ur Theoretische Physik, 
        Universit\"at Heidelberg\\
        E-mail: \email{plehn@uni-heidelberg.de}}

\author{Heidi Rzehak\\
        Institut f\"ur Theoretische Physik, 
        Karlsruhe Institute of Technology (KIT)\\
        E-mail: \email{hr@particle.uni-karlsruhe.de}}

\abstract{We present a full calculation of the supersymmetric NLO
corrections to Higgs boson production via vector-boson fusion. The
supersymmetric QCD corrections turn out to be significantly smaller than
the electroweak ones. These higher-order corrections are an important
ingredient to a precision analysis of the Higgs sector at the LHC.  }

\FullConference{RADCOR 2009 - 9th International Symposium on Radiative Corrections (Applications of Quantum Field Theory to Phenomenology) \\
		 October 25-30 2009\\
		 Ascona, Switzerland}

\begin{document}

\section{Introduction}

Understanding the origin of electroweak symmetry breaking is one of the
main tasks of the Large Hadron Collider (LHC). Electroweak precision
data predicts a light Higgs boson. To solve the hierarchy problem the
Standard Model (SM) needs to be embedded in a larger theory for an
ultra-violet completion. The Minimal Supersymmetric Standard Model
(MSSM) is a very promising candidate for that. Higgs-boson production in
vector-boson fusion with a subsequent decay of the Higgs into a pair of
tau leptons is one of the most auspicious channels for an early
discovery~\cite{wbf_ex,wbf_tau}. The discovery reach for this channel
covers the entire MSSM parameter space~\cite{wbf_susy}.

Determining the relations in the Higgs sector, like the gauge and
Yukawa couplings~\cite{duhrssen}, will then be the next step for the
LHC. These measurements require a good knowledge of the associated rates
and their theory errors need to be under control, including 
higher-order effects. The next-to-leading-order QCD corrections to
vector-boson-fusion Higgs production are fairly small, of the order of
ten percent~\cite{wbf_nloqcd}. This is due to the color structure of the
process which forbids gluon exchange between the two quark lines,
combined with its forward-jet nature. The electroweak corrections turn
out to be of similar size and have opposite sign for the
phenomenologically relevant region of a light Higgs
boson~\cite{wbf_nloew}. Also NNLO-QCD
effects~\cite{NNLO} and the interference between vector-boson-fusion and
gluon-fusion Higgs production~\cite{gVBF_interf} have been investigated
and found to be tiny.

For the MSSM these calculations must be augmented by the corrections
originating from the additional supersymmetric particles and the
extended Higgs sector. They need to be included for the determination of
the Higgs sector, either as correction or as a theory uncertainty for
early running or if the MSSM spectrum is not favorable to precision MSSM
analyses~\cite{sfitter}. Both cases need a detailed study of the
supersymmetric contributions to vector-boson
fusion~\cite{susynloold,susynlo} and
also gluon-fusion Higgs production~\cite{maggi}.

\section{Corrections in the MSSM Higgs sector}

The coupling of the light supersymmetric Higgs to vector bosons receives
an additional factor $\sin(\beta-\alpha)$ compared to its SM counter
part, where $\tan\beta$ is the ratio of the vacuum expectation values of
the two Higgs doublets and $\alpha$ the mixing angle turning the two
CP-even degrees of freedom into mass eigenstates. Hence for a given Higgs
mass we obtain the tree-level MSSM cross section by simply rescaling the
SM rate with $\sin^2(\beta-\alpha)$. For pseudoscalar Higgs masses $m_A$
exceeding $200~\gev$ this factor is close to one.
At next-to-leading order two additional types of diagrams appear: First,
there are the ones where the SM particles in the loop are replaced by
their supersymmetric counterparts. As we assume $R$ parity conservation,
a loop consists purely of either SM or supersymmetric particles. This
allows for a diagrammatic separation of the new MSSM contributions.
Second, contributions from the extended Higgs sector of the MSSM appear.
As there is no one-to-one correspondence of the Standard-Model Higgs to
an MSSM particle, we cannot separate them at the diagram level. Instead,
we compute the full MSSM-Higgs corrections at the amplitude level and
then subtract the SM Higgs part, multiplied by the tree-level
correction factor. Since both the full MSSM and the SM are
gauge-invariant, our supersymmetric corrections calculated in that way
share this feature.

The large number of diagrams requires us to use automated tools for the
evaluation. We compute the cross section with HadCalc~\cite{hadcalc}
using the MRST 2002 NLO pdf set~\cite{mrst2002}, where we have generated
the Feynman diagrams and amplitudes with FeynArts and
FormCalc~\cite{feynarts}. The evaluation of the loop integrals we
perform with LoopTools~\cite{looptools}. For an optimal mapping of the
phase-space integration we use code from VBFNLO~\cite{vbfnlo}.  We
assume minimal flavor violation, whose effect is small after taking all
experimental and theoretical constraints into account~\cite{nmfv}, as
well as a $CP$-conserving MSSM.

The mass of the lightest Higgs boson receives large loop
corrections~\cite{mh_feyn, feynhiggs, mh_eff, subh}, which push its
value beyond the limits from LEP2~\cite{lep2}. Therefore, we need to
include them for a phenomenologically relevant analysis. Linked to the
mass shift are corrections to the Higgs couplings~\cite{higgsmc}, which
we should add at the same order in perturbation theory. For the
numerical evaluation of the Higgs sector we use FeynHiggs
2.6.2~\cite{feynhiggs}. We have checked that the differences to a
program using the effective-potential approach~\cite{subh} are
small~\cite{susynlo}. 

Consistent with FeynHiggs, we perform the renormalization of the Higgs
sector including $\tan\beta$ in the $\drbar$ scheme. According to the
LSZ prescription we then need to add finite wave-function
renormalization terms to ensure that the residue of the Higgs boson pole
is unity. We include them as an additional one-loop correction with the
amplitude   
\begin{equation}
\Gamma = (\sqrt{Z_{hh}} - 1) \Gamma_{h_0} + 
          \sqrt{Z_{hh}} Z_{hH} \Gamma_{H_0} \ ,
\end{equation}
where $\Gamma_{h_0}$ and $\Gamma_{H_0}$ are the tree-level amplitudes
for vector-boson-fusion $h^0$ and $H^0$ production, respectively. We use 
\begin{eqnarray}
\sqrt{Z_{hh}} = 1 - \frac12 \left. \mRe \left( \frac{\di}{\di p^2} 
                \hat\Sigma_{hh}( p^2 ) \right)
                 \right|_{p^2=m_{h^0}^2} \\
Z_{hH} = \frac1{m_{H^0}^2 - m_{h^0}^2} 
           \mRe ( \hat\Sigma_{hH}(m_{h^0} )) 
\end{eqnarray}
where the $\hat\Sigma$ are the renormalized one-loop self
energies~\cite{feynhiggs} and the Higgs boson masses are 
loop-corrected. All Standard-Model parameters are
renormalized on-shell.

For our numerical analysis we require the following standard
vector-boson-fusion cuts:
\begin{eqnarray}
y_j < 4.5 \ , \qquad (p_T)_j > 20~\gev \ , \qquad
  y_{j_1} \cdot y_{j_2} < 0 \ , \nonumber \\
  |y_{j_1}-y_{j_2}| > 4.5 \ , \qquad
  m_{\mathrm{inv}}(j_1,j_2) > 600~\gev \ .
\end{eqnarray}

\section{Supersymmetric corrections}
\begin{figure*}[t]
\includegraphics[width=0.18\textwidth]{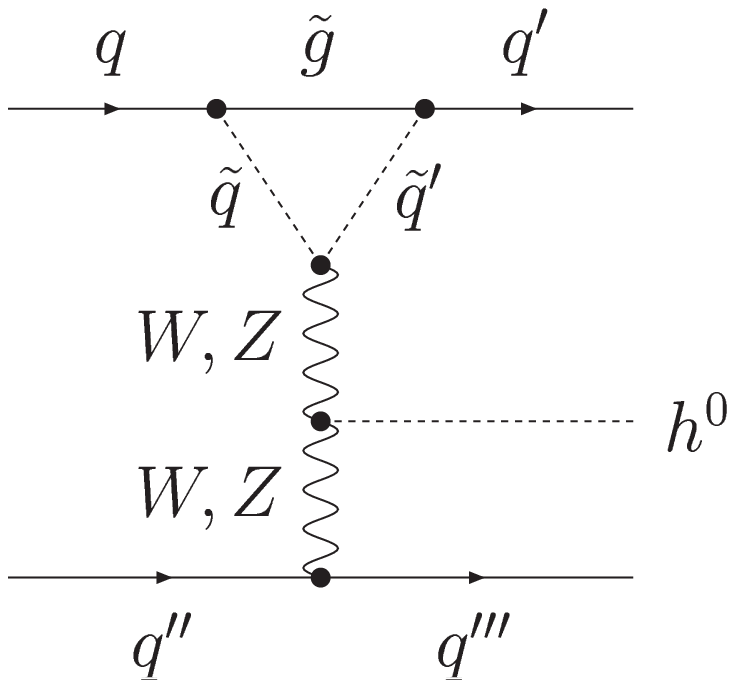}
\hspace*{5mm}
\includegraphics[width=0.20\textwidth]{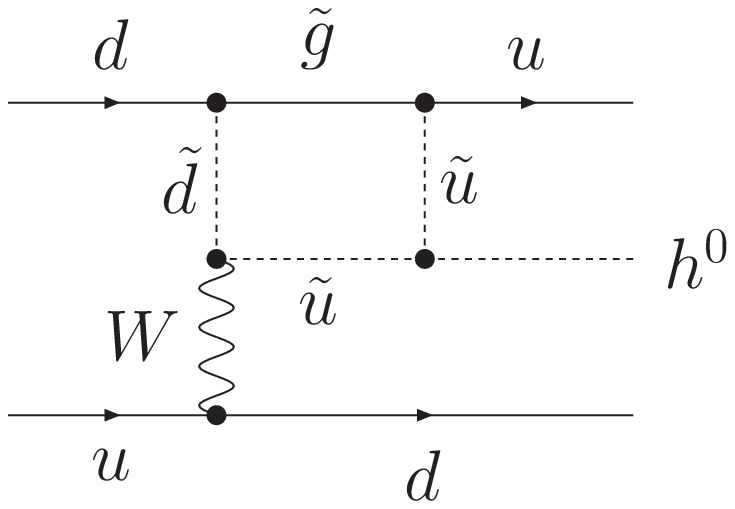}
\hspace*{5mm}
\includegraphics[width=0.20\textwidth]{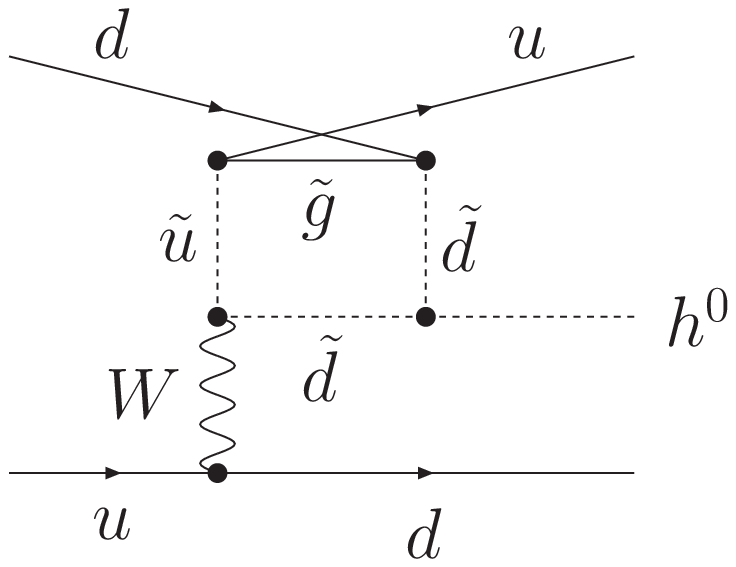}
\hspace*{5mm}
\includegraphics[width=0.22\textwidth]{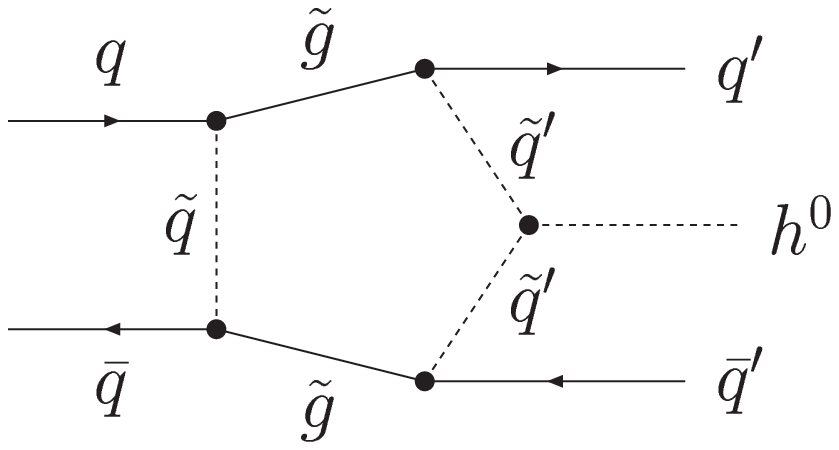} \\[2mm]
\includegraphics[width=0.18\textwidth]{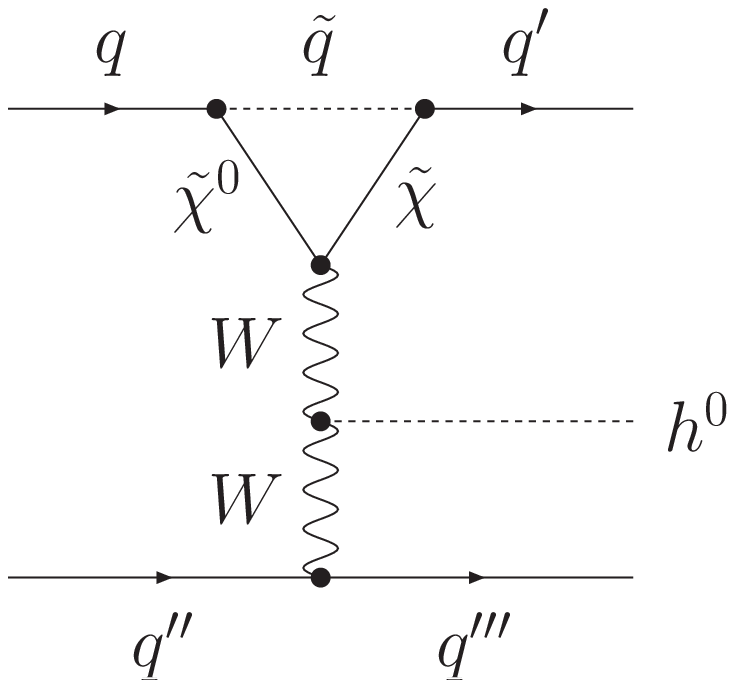}
\hspace*{-1mm}
\includegraphics[width=0.20\textwidth]{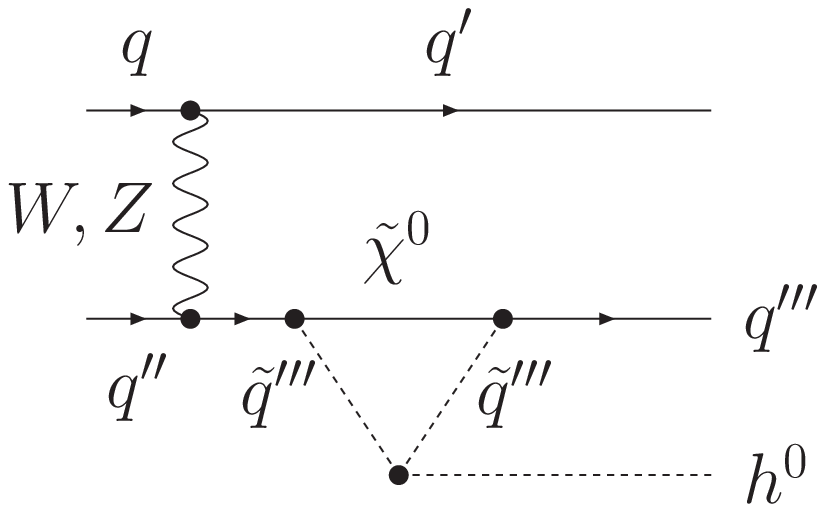}
\hspace*{-1mm}
\includegraphics[width=0.16\textwidth]{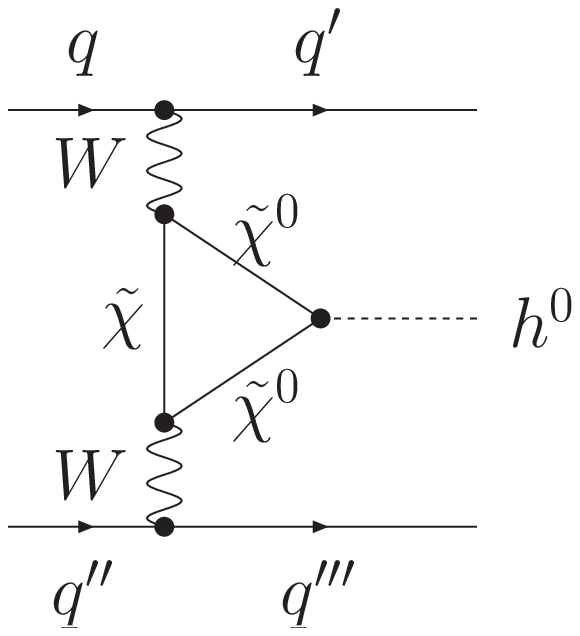}
\hspace*{-1mm}
\includegraphics[width=0.20\textwidth]{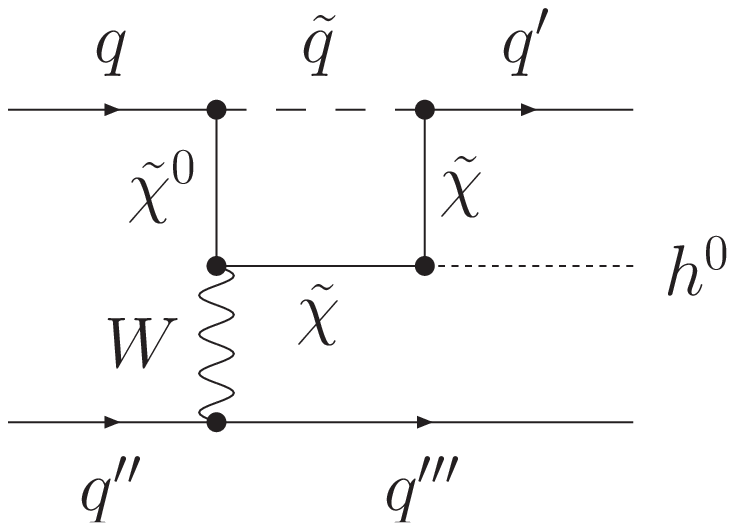}
\hspace*{-1mm}
\includegraphics[width=0.22\textwidth]{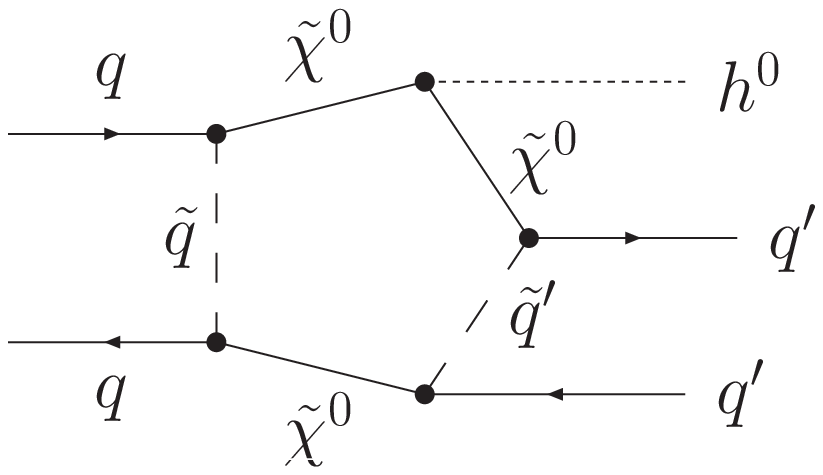}
\caption{Example Feynman diagrams contributing
  to QCD (upper) and electroweak (lower) vertex corrections, 
  boxes and pentagons.}
\label{fig:feyn}
\end{figure*}

In Fig.~\ref{fig:feyn} we show an example set of one-loop Feynman
diagrams which appear in our calculation. We also include
diagrams where the Higgs is radiated off the quark lines or the
weak bosons are replaced by photons.
The numerical results we first show for the parameter point
SPS1a~\cite{sps}, where we have used the high-scale definition and
evolved them to the low scale using SoftSUSY 2.0.14~\cite{softsusy}.
The electromagnetic coupling constant we choose at vanishing momentum.
Also we neglect all masses of the first two quark and lepton
generations.  The spectrum of this parameter point consists of fairly
light particles with masses of uncolored particles typically at
$100-200~\gev$ and squarks and gluinos around $500-600~\gev$, and
$\tan\beta = 10$ leading to only small decoupling effects in the
down-sector.
We have compared our results to Ref.~\cite{susynloold}, where the
vertex-correction contributions of the upper-left diagram of
Fig.~\ref{fig:feyn} have already been evaluated in the limit of equal
squark masses, as well as to an upcoming second calculation of these
corrections~\cite{sophy}.
In both cases we find good agreement.

\begin{table}
\begin{tabular}{l||r|r}
\hline
diagram & $\Delta \sigma/\sigma~[\%]$ & $\Delta \sigma/\sigma~[\%]$ \\
\hline
  & $\Delta \sigma \sim \mathcal{O}(\alpha)$  
  & $\Delta \sigma \sim \mathcal{O}(\alpha_s)$ \\
\hline
self energies      &  0.199    & \\
\hline
$qq'W+qqZ$         & -0.392    & -0.0148      \\
$qqh$              & -0.0260   &  0.00545     \\
$WWh+ZZh$          & -0.329    & \\
\hline
box                &  0.0785   & -0.00518     \\
pentagon           &  0.000522 & -0.000308    \\ 
\hline
\multicolumn{3}{c}{sum of all $\Delta \sigma/\sigma = -0.484~\%$}
\end{tabular}
\hfill
\begin{tabular}{l||r|r|r||r}
\multicolumn{5}{c}{$\Delta \sigma/\sigma~[\%] $} \\
\hline
& $VVh$
& $\mathcal{O}(\alpha)$
& $\mathcal{O}(\alpha_s)$
& all \\
\hline
SPS1a & -0.329 & -0.469 & -0.015 & -0.484 \\
SPS1b & -0.162 & -0.229 & -0.006 & -0.235 \\
SPS2  & -0.147 & -0.129 & -0.002 & -0.131 \\
SPS3  & -0.146 & -0.216 & -0.006 & -0.222 \\
SPS4  & -0.258 & -0.355 & -0.008 & -0.363 \\
SPS5  & -0.606 & -0.912 & -0.010 & -0.922 \\
SPS6  & -0.226 & -0.309 & -0.010 & -0.319 \\ 
SPS7  & -0.206 & -0.317 & -0.006 & -0.323 \\
SPS8  & -0.157 & -0.206 & -0.004 & -0.210 \\
SPS9  & -0.094 & -0.071 & -0.003 & -0.074 
\end{tabular}
\caption{Complete MSSM corrections to the process $pp \to qqh$
by diagram types for the parameter point SPS1a (left) and for all SPS
points (right). Tables taken from Ref.~\cite{susynlo}.}
\label{tab:sps}
\end{table}

All supersymmetric QCD corrections turn out to be very small as we can
see in Table~\ref{tab:sps}. We find an even bigger suppression than in
the Standard Model, where gluon exchange between the two quark lines
leads to a vanishing color trace. 

Two tree-level vertices receive one-loop corrections, $qqV$ and $VVh$.
Only at the first one squark/gluino loops appear. Since the $W$ boson
couples only to left-handed particles and the mixing between left- and
right-handed light-flavor squarks is negligible, like at tree-level both
external quarks are then left-handed. Therefore, the gluino propagator in
the fermion trace can only contribute via its momentum and not via a
gluino-mass insertion which would require a chirality flip. Hence the
typical scale in the numerator is $m_h/2$, an order of magnitude below
the gluino mass in the denominator.

In the electroweak case, also the lighter charginos and neutralinos in
the loop couple to the vector boson. This means that we can add a double
mass insertion into the fermion line which can partly compensate for the
heavy masses in the loop denominator. This effect leads to a relative
enhancement of the electroweak over the QCD $qqV$ vertex correction we
observe in Table~\ref{tab:sps}.

In both box diagrams shown in the upper line of Fig.~\ref{fig:feyn} the
$\tilde{q}\tilde{q}'W$ and $q\tilde{q}\tilde{g}$ couplings are the same,
while the $\tilde{q}\tilde{q}h$ coupling is proportional to $T_3 - Q
s_W^2$, which is around $\frac13$ for the up and $-\frac5{12}$ for
the down sector. Therefore a cancellation of roughly one order of
magnitude occurs. This cannot be broken by different squark
masses, because the left-handed squarks form an $SU(2)$ doublet and again
left-/right-handed squark mixing effects are tiny. For the subleading
$ZZ$ fusion channel and the electroweak corrections this argument does
not hold. Here we indeed find corrections at a more natural level.

For supersymmetric pentagon diagrams there is an additional possibility
with two colored particles exchanged, depicted in the upper-right corner
of Fig.~\ref{fig:feyn}. Formally, the interference with the tree-level
diagram is of order $\om(\alpha_s^2\alpha^2)$, which is as large as the
Born contribution $\om(\alpha^3)$. However, these diagrams have
completely different kinematic properties compared to the
vector-boson-fusion topology. Combined with the large loop masses this
leads to a negligible contribution.

From these arguments we see that there is not a single explanation for
the smallness of the supersymmetric QCD corrections, but a set of
mechanisms which can explain these at first sight surprising results.

On the right-hand side of Table~\ref{tab:sps} we show numerical results
for all SPS parameter points, which probe typical different parts of the
MSSM parameter space. As expected, the overall picture of our numerical
results stays unchanged. We see that the supersymmetric QCD corrections
are strongly suppressed and their electroweak counterpart is less or
around one percent. A large mass splitting in the stop sector leads to
comparably large effects for the SPS5 point.

\section{Conclusions}

We have presented a calculation of the next-to-leading order
supersymmetric corrections to Higgs-boson production via
vector-boson-fusion in the MSSM. This is an important ingredient for a
precision analysis of the Standard-Model and the MSSM Higgs sector at
the LHC. We find that the supersymmetric QCD corrections are reduced to
a negligible level. This is due to various effects ranging from the
color structure and the coupling structure to the kinematics of the
process. The supersymmetric $\om(\alpha)$ contributions turn out to be at
the percent level and therefore at a typical size for massive electroweak
corrections. 
In general, the total corrections can reach up to four percent for
parameter points still allowed by direct searches, with usual sizes at
or below one percent, and typically negative sign.

\end{document}